Derrick, B., White, P., and Toher, D.

# An Inverse Normal Transformation Solution for the comparison of two samples that contain both paired observations and independent observations.

## Abstract

Inverse normal transformations applied to the partially overlapping samples t-tests by Derrick *et.al.* (2017) are considered for their Type I error robustness and power. The inverse normal transformation solutions proposed in this paper are shown to maintain Type I error robustness. For increasing degrees of skewness they also offer improved power relative to the parametric partially overlapping samples t-tests. The power when using inverse normal transformation solutions are comparable to rank based non-parametric solutions.

## Introduction

A frequently asked question in research is how to compare means between two samples that include both paired observations and unpaired observations (Derrick, Toher and White, 2017). This is referred to as 'partially overlapping samples' (Derrick *et.al.*, 2015).

Parametric test statistics for the comparison of means of two samples that contain both paired observations and independent observations are given by Derrick *et.al.* (2017). These partially overlapping samples t-tests use all of the available data, and are Type I error robust and more powerful than conventional methods, when any missing data is missing completely at random.

The partially overlapping samples t-tests are based on the assumption of normality (Derrick, 2017). To remove the restriction of the normality assumption, the application of an Inverse Normal Transformation (INT) is proposed in this paper. This method is compared to the existing partially overlapping samples t-tests, and a simple ranking based non-parametric method.

The parametric partially overlapping samples test statistic, $T_{new1}$, is an interpolation between the paired samples t-test and the independent samples t-test assuming equal variances, and is given by Derrick *et.al.* (2017) as:

$$T_{new1} = \frac{\overline{X}_1 - \overline{X}_2}{S_p\sqrt{\frac{1}{n_1} + \frac{1}{n_2} - 2r\left(\frac{n_c}{n_1 n_2}\right)}} \quad \text{where} \quad S_p = \sqrt{\frac{(n_1-1)S_1^2 + (n_2-1)S_2^2}{(n_1-1)+(n_2-1)}}$$

The test statistic $T_{new1}$ is referenced against the t-distribution with degrees of freedom:

$$v_1 = (n_c - 1) + \left(\frac{n_a + n_b + n_c - 1}{n_a + n_b + 2n_c}\right)(n_a + n_b).$$

where for j = {Group 1, Group 2}, $\overline{X}_j$ = mean of Sample j, $n_a$ = number of unpaired observations exclusive to Sample 1, $n_b$ = number of unpaired observations exclusive to Sample 2, $n_c$ = number of pairs, $n_j$ = total number of observations in Sample j, $S_j^2$ = variance of Sample j, and $r$ = Pearson's correlation coefficient for the paired observations.

The independent samples t-test is not Type I error robust when variances are not equal and sample sizes are not equal. It follows that $T_{new1}$ is also sensitive to deviations from the equal variances assumption. If equal variances cannot be assumed, then alternatively Welch's t-test is Type I error robust under normality (Derrick, Toher and White, 2016). The partially overlapping samples t-test not constrained to equal variances, $T_{new2}$, is an interpolation between the paired samples t-test and Welch's t-test, is given by Derrick *et.al.* (2017) as:

Derrick, B., White, P., and Toher, D.

$$T_{new2} = \frac{\overline{X}_1 - \overline{X}_2}{\sqrt{\frac{S_1^2}{n_1} + \frac{S_2^2}{n_2} - 2r\left(\frac{S_1 S_2 n_c}{n_1 n_2}\right)}}$$

The test statistic $T_{new2}$ is referenced against the t-distribution with degrees of freedom:

$$v_2 = (n_c - 1) + \left(\frac{\gamma - n_c + 1}{n_a + n_b + 2n_c}\right)(n_a + n_b) \text{ where } \gamma = \frac{\left(\frac{S_1^2}{n_1} + \frac{S_2^2}{n_2}\right)^2}{\frac{(S_1^2/n_1)^2}{n_1 - 1} + \frac{(S_2^2/n_2)^2}{n_2 - 1}}$$

For simple ranking based non-parametric solutions, observations are pooled and assigned rank values in ascending order. The rank values are substituted into the elements of the calculation for $T_{new1}$ and $T_{new2}$ in place of the observed values. This gives the test statistics $T_{RNK1}$ and $T_{RNK2}$ respectively. The degrees of freedom $v_1$ and $v_2$ respectively, are calculated using the pooled rank values.

The transforming of data attempts to overcome violations of the normality assumption, so that the traditional parametric tests can be applied. For equal sample sizes, Cohen and Arthur (1991) found that the independent samples t-tests performed on log or squared data, exhibits satisfactory Type I error robustness. However, popular transformations such as the Box-Cox transformations do not necessarily lead to Type I error robust tests when sample sizes are unequal or variances are unequal (Zarembka, 1974). In addition the Box-Cox transformation rarely results in both normality and equal variances at the same time (Sakia, 1992). Even if researchers are comfortable with the hypothesis of comparing means of transformed data, a suitable transformation may not always be found.

A transformation that should always give the appearance of normally distributed data, is an Inverse Normal Transformation (INT), which derives properties from the Normal distribution. Methods based on Fisher and Yates Normal approximation are the most commonly used INT (Beasley and Erickson, 2009), and are the most

Derrick, B., White, P., and Toher, D.

powerful (Bradley, 1968). Example INT approaches include Blom's method and Van der Waerden's method, but it is of little consequence which of them is selected because they are linear transformations of one-another (Tukey, 1962). It should be noted that an INT does not make a population Normal, but it makes a sample appear Normal. This is not the same as directly ensuring the assumption of normally distributed residuals is true (Servin and Stephens, 2007). In addition, care needs to be taken with the interpretation of results. The null hypothesis of equal means for the transformed data, is assumed to be equivalent to a null hypothesis of equal means.

The Fisher and Yates INT procedure requires pooling all of the observations, sorting into ascending order and ranking each observation over the entire sample so that:

$X_i = \Phi^{-1}\left(\dfrac{y_i - c}{N - 2c + 1}\right)$ where $y_i$ is the ordinary rank of observation $i$, $N$ is the total pooled sample size, and $\Phi^{-1}$ is the standard Normal quantile function. The most simple INT method is attributed to Van Der Waerden (1952) uses $c = 0$. Penfield (1994) found that Van Der Waerden transformations applied to the independent samples t-test, are Type I errors robust across a variety of distributions.

Calculating the Van Der Waerden scores, $X_m$, and using these within the calculation of $T_{new1}$ and $T_{new2}$, gives distribution free test statistics $T_{INT1}$ and $T_{INT2}$ respectively. The degrees of freedom $\upsilon_1$ and $\upsilon_2$ respectively, are calculated using the pooled transformed values. The calculation of *r* is Pearson's correlation coefficient between the transformed paired observations.

## Methodology

For each of the six test statistics defined above ($T_{new1}$, $T_{new2}$, $T_{RNK1}$, $T_{RNK2}$, $T_{INT1}$, $T_{INT2}$), the robustness for validity (i.e. Type I errors) and efficacy (i.e. power) are explored using simulation. Values of $n_a$, $n_b$, $n_c$ {5, 10, 30, 50, 100, 500} are varied in a factorial design, as well as values of the population correlation coefficient $\rho$ {-.75, -.50, -.25, .00, .25, .50, .75}.

To generate the independent observations, firstly the Mersenne-Twister algorithm (Matsumoto and Nishimura, 1998) generates random U(0,1) deviates. These uniform deviates are transformed into N(0,1) deviates using the Paley and Wiener (1934) transformation.

To generate the paired observations, the approach used is equivalent to that used by Derrick and White (2017) and Derrick *et.al.* (2017). In this approach, additional Standard Normal deviates are transformed to correlated Standard Normal bivariates, $z_{ji}$, as follows:

$$z_{1i} = \sqrt{\frac{1+\rho}{2}}z_{1i} + \sqrt{\frac{1-\rho}{2}}z_{2i} \text{ and } z_{2i} = \sqrt{\frac{1+\rho}{2}}z_{1i} - \sqrt{\frac{1-\rho}{2}}z_{2i} \text{ where } i = (1, 2, \ldots, n_c)$$

Each of the test statistics are assessed firstly under the Standard Normal distribution. For the comparison of test statistics under non-normality, observations are generated by transformation of the Standard Normal deviates as given in Table 1.

Table 1. Transformations applied to Standard Normal deviates (N) to obtain non-normally distributed deviates (X), with the resulting skewness and kurtosis. Note that Uniform (U) deviates are calculated as the cumulative density function of N.

| Distribution | Transformation | Skewness | Kurtosis |
|---|---|---|---|
| Normal (N) | X | 0.000 | 3.000 |
| Gumbel | X= -log(-log U) | 1.140 | 5.400 |
| Exponential | X= -log (U) -1 | 2.004 | 9.000 |
| Lognormal | X= exponential (N) | 6.145 | 107.256 |

Derrick, B., White, P., and Toher, D.

Following the transformations as per Table 1, the calculation of ranks / inverse normal transformations, are performed as appropriate for each statistical test. Each of the tests are performed at the 5% significance level, two sided.

For each of the parameter combinations within the factorial design, the null hypothesis rejection rate (NHRR) is recorded as the proportion of the 10,000 replicates where the null hypothesis is rejected. The methodology is depicted in Figure 1. Analyses under the alternative hypothesis proceeds similarly, but with the addition of 0.5 to the $n_2$ observations directly following the transformation in Table 1.

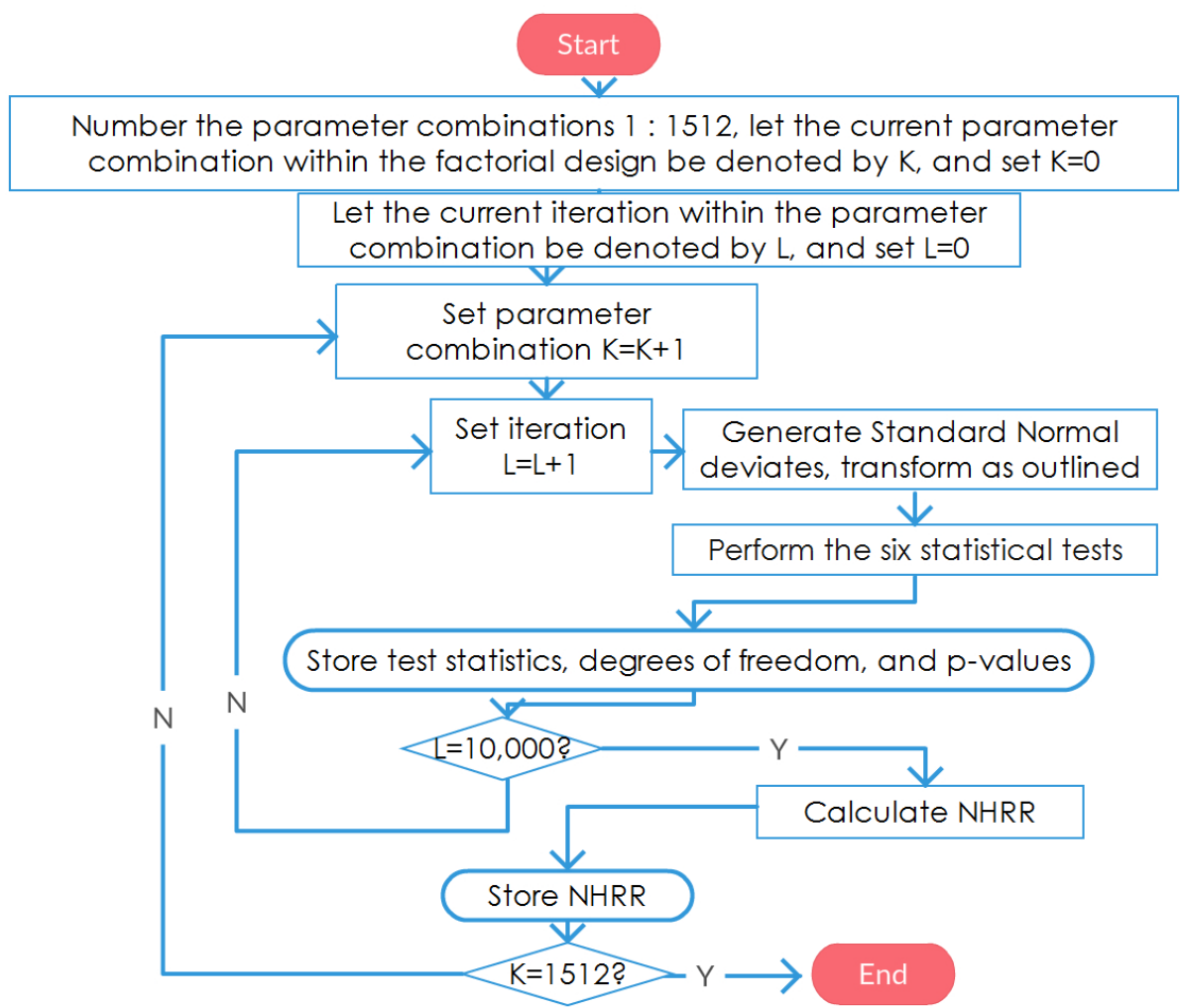

Figure 1. Overview of the simulation process.

# Results

Under the null hypothesis, 10,000 iterations are obtained for each of the 1,512 parameter combinations. The Type I error rates for each of the test statistics across the simulation design are given through Figure 2 to Figure 5. Each parameter combination which has a Type I error rate between 2.5% and 7.5%, is considered to be maintaining reasonable Type I error robustness

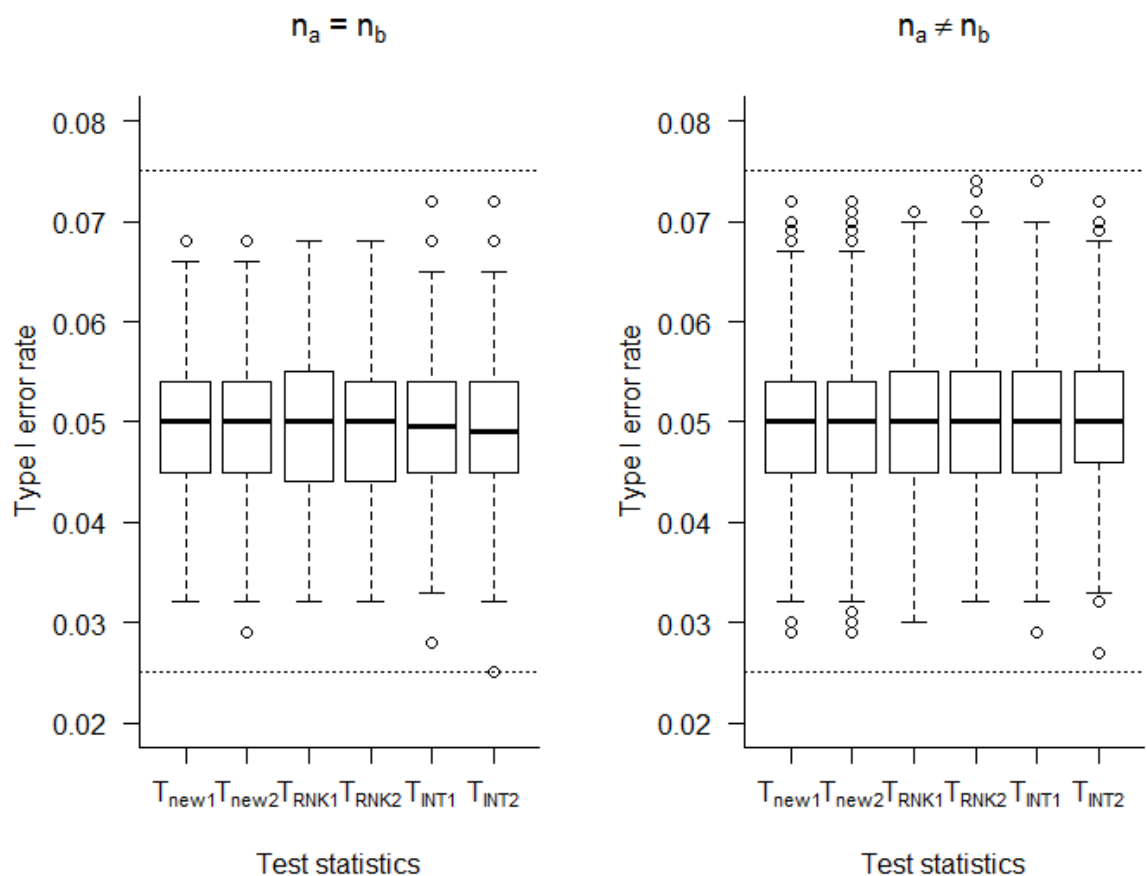

Figure 2. Type I error rates for the Standard Normal distribution.

Figure 2 shows that when both samples are taken from the Standard Normal distribution, all of the test statistics are Type I error robust for all of the parameter combinations within the simulation design.

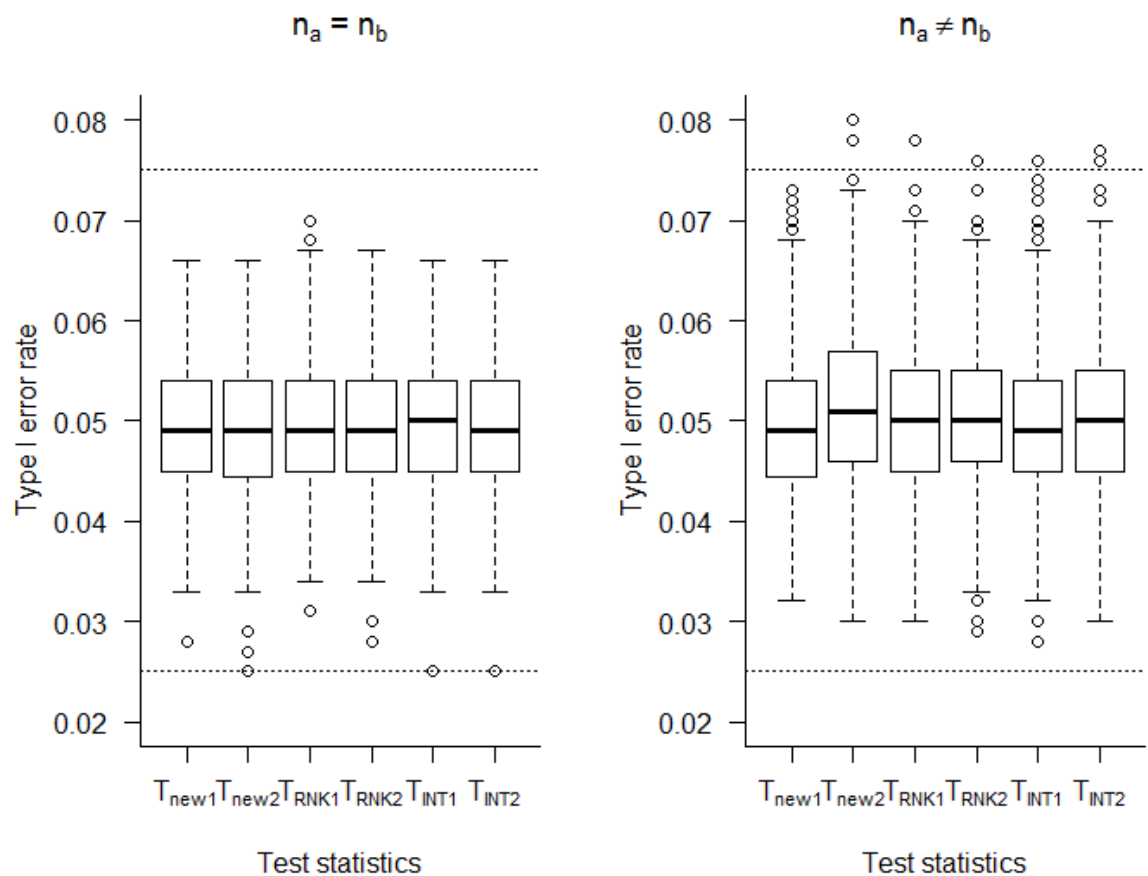

Figure 3. Type I error rates for the Gumbel distribution.

Figure 3 shows that when both samples are taken from the Gumbel distribution, all of the test statistics are Type I error robust for all of the parameter combinations within the simulation design. This suggests that all of the test statistics are Type I error robust for distributions with a relatively small degree of skewness.

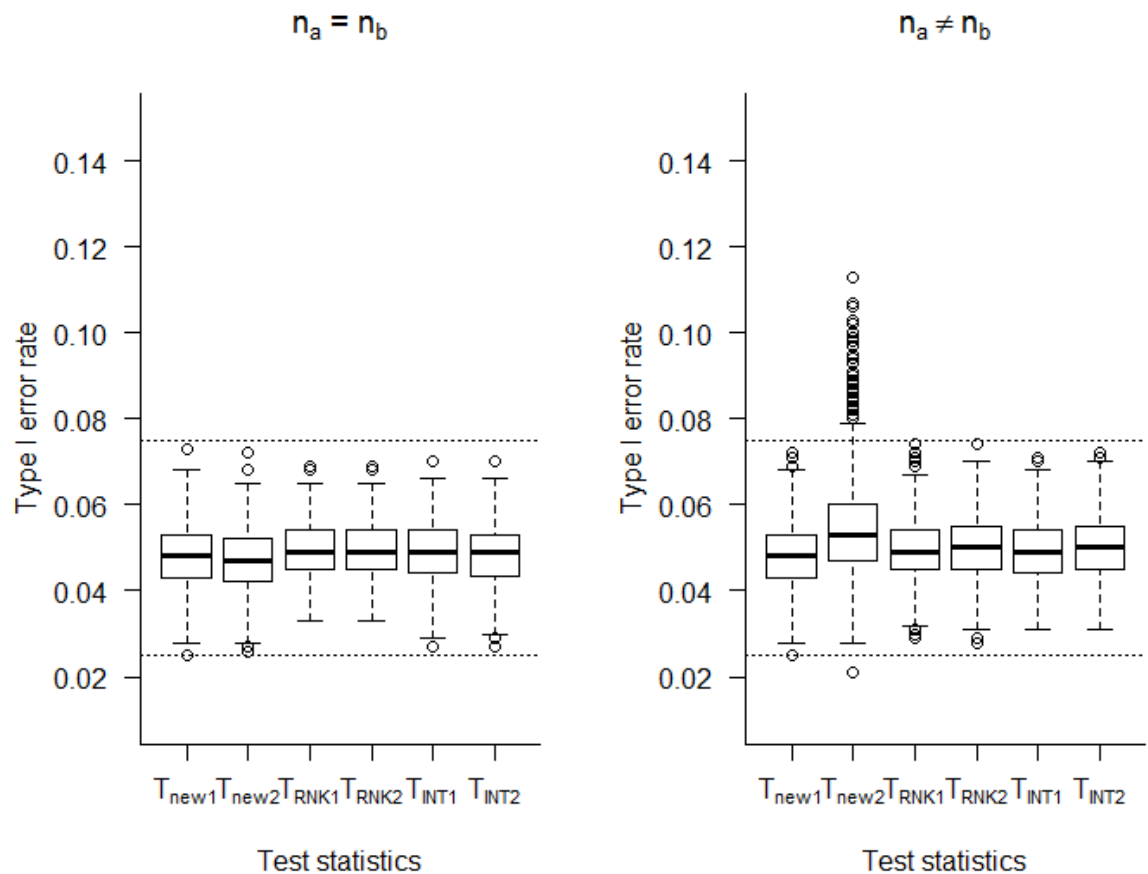

Figure 4. Type I error rates for the Exponential distribution.

Figure 4 shows that when both samples are taken from the Exponential distribution, all of the test statistics are Type I error robust when sample sizes are equal. However the test statistic $T_{new2}$ is not Type I error robust when there is a large imbalance between the size of the two samples. This suggests that all of the test statistics except $T_{new2}$ are Type I error robust for distributions with a moderate skew.

Derrick, B., White, P., and Toher, D.

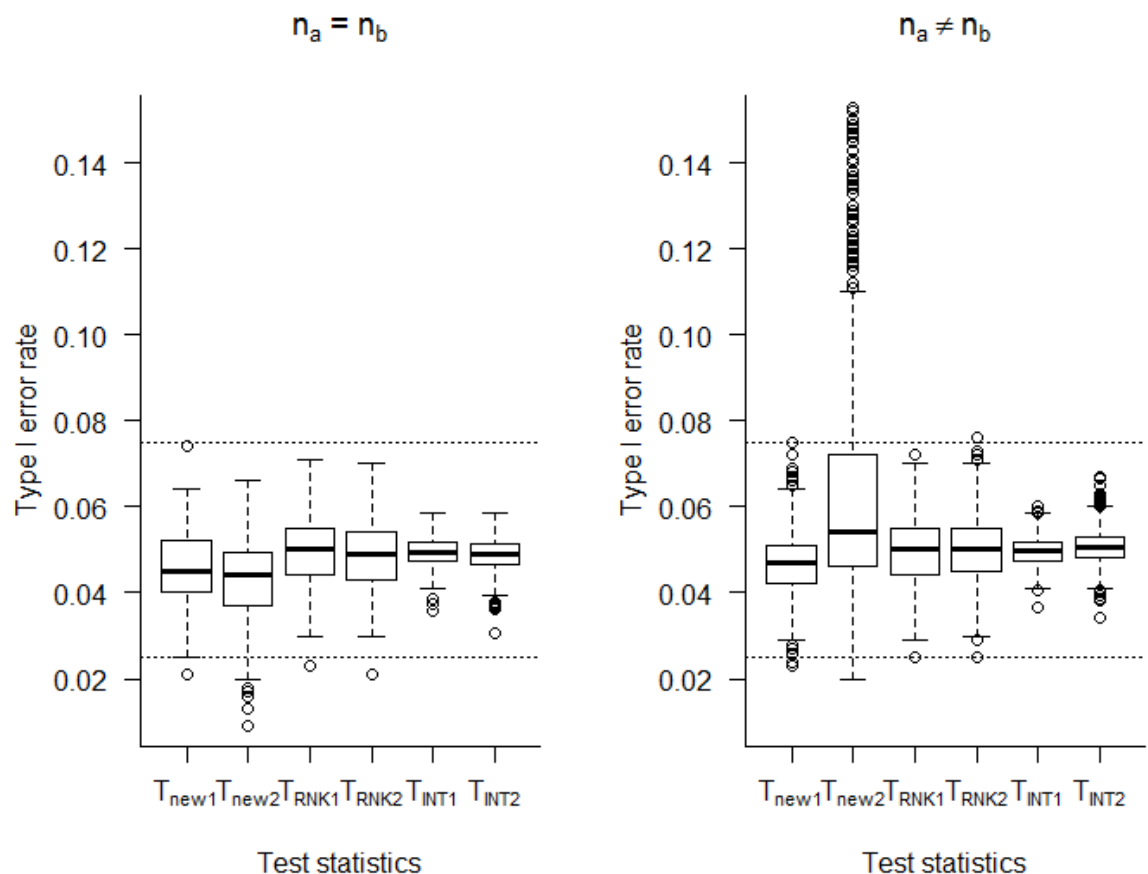

Figure 5. Type I error rates for the Lognormal distribution.

Figure 5 shows that when both samples are taken from the Lognormal distribution, all of the test statistics are Type I error robust with the exception of $T_{new2}$ which can be occasionally conservative when sample sizes are equal, and can be liberal when sample sizes are not equal. This suggests that all of the test statistics except $T_{new2}$ maintain Type I error robustness for heavily skewed distributions.

For the simulations under the alternative hypothesis Table 2 gives the average power across the simulation design of sample size and correlation coefficient for each of the distributions. Power is only recorded for scenarios where the test statistic maintains Type I error robustness.

Table 2. Power for $\alpha = .05$, $\mu_2 - \mu_1 = .5$, two sided, over all values of $n_c$.

| Distribution | Sample size | $\rho$ | $T_{new1}$ | $T_{new2}$ | $T_{RNK1}$ | $T_{RNK2}$ | $T_{INT1}$ | $T_{INT2}$ |
|---|---|---|---|---|---|---|---|---|
| Normal | $n_a = n_b$ | >0 | .865 | .864 | .856 | .855 | .855 | .854 |
| | | 0 | .819 | .819 | .811 | .811 | .811 | .811 |
| | | <0 | .779 | .779 | .772 | .771 | .770 | .769 |
| | $n_a \neq n_b$ | >0 | .839 | .832 | .829 | .824 | .827 | .824 |
| | | 0 | .806 | .798 | .795 | .790 | .795 | .790 |
| | | <0 | .774 | .767 | .763 | .760 | .761 | .758 |
| Gumbel | $n_a = n_b$ | >0 | .783 | .782 | .815 | .814 | .824 | .823 |
| | | 0 | .720 | .718 | .761 | .760 | .774 | .774 |
| | | <0 | .678 | .678 | .719 | .719 | .727 | .726 |
| | $n_a \neq n_b$ | >0 | .740 | .735 | .779 | .776 | .789 | .786 |
| | | 0 | .693 | .689 | .740 | .736 | .749 | .747 |
| | | <0 | .655 | .651 | .702 | .699 | .712 | .710 |
| Exponential | $n_a = n_b$ | >0 | .867 | .864 | .938 | .937 | .946 | .944 |
| | | 0 | .824 | .824 | .915 | .914 | .926 | .925 |
| | | <0 | .795 | .795 | .894 | .894 | .906 | .906 |
| | $n_a \neq n_b$ | >0 | .841 | | .933 | .930 | .943 | .938 |
| | | 0 | .811 | - | .919 | .917 | .930 | .926 |
| | | <0 | .786 | | .904 | .903 | .918 | .915 |
| Lognormal | $n_a = n_b$ | >0 | .596 | .590 | .893 | .891 | .905 | .904 |
| | | 0 | .535 | .533 | .857 | .856 | .911 | .912 |
| | | <0 | .506 | .506 | .826 | .826 | .918 | .925 |
| | $n_a \neq n_b$ | >0 | .514 | | .874 | .873 | .879 | .876 |
| | | 0 | .467 | - | .851 | .850 | .850 | .851 |
| | | <0 | .438 | | .825 | .826 | .848 | .849 |

When population variances are equal, Table 2 shows that the test statistics not assuming equal variances, $T_{new2}$, $T_{RNK2}$ and $T_{INT2}$, perform similarly to their counterparts where equal variances are assumed $T_{new1}$, $T_{RNK1}$ and $T_{INT1}$ respectively.

From Table 2 it can be seen that for normally distributed data, the parametric statistics $T_{new1}$ and $T_{new2}$ are marginally more powerful than the other test statistics considered, but not to any meaningful extent.

The relative power advantage of $T_{INT1}$ and $T_{INT2}$ over $T_{new1}$ and $T_{new2}$ increases as the degree of skewness increases. However, the proposed statistics using inverse normal transformations, $T_{INT1}$ and $T_{INT2}$, yield very similar results to $T_{RNK1}$ and $T_{RNK2}$.

# Conclusion

Test statistics making use of all of the available data in a partially overlapping samples design are compared using simulation. Assuming normality the partially overlapping samples t-test proposed by Derrick *et.al.* (2017) for equal variances, $T_{new1}$, is more powerful than non-parametric equivalents and Inverse Normal transformations.

The test statistics making use of Inverse Normal Transformations offer no substantial improvement over the non-parametric tests. Due to its Type I error robustness, power properties and relative simplicity, $T_{RNK1}$ is recommended over $T_{INT1}$ as the best solution for comparing partially overlapping samples from non-normal distributions.

# References

Beasley, T. M., Erickson, S. & Allison, D. B. (2009). Rank-based inverse normal transformations are increasingly used, but are they merited?. Behavior genetics. 39(5), 580-595.

Bradley, J. V. (1968). Distribution-free statistical tests. Prentice-Hall; New York.

Cohen, M. & Arthur, J. (1991). Randomization analysis of dental data characterized by skew and variance heterogeneity. Community dentistry and oral epidemiology. 19(4), 185-189.

Derrick, B. (2017) Statistics: New t-tests for the comparison of two partially overlapping samples. In: Faculty of Environment and Technology Degree Show, UWE, Frenchay Campus, UWE, 1 June 2017. Available from: http://eprints.uwe.ac.uk/31766.

Derrick, B., Dobson-McKittrick, A., Toher, D. & White P. (2015). Test statistics for comparing two proportions with partially overlapping samples. Journal of Applied Quantitative Methods. 10(3), 1-14.

Derrick, B., Russ, B., Toher, D. & White P. (2017). Test statistics for the comparison of means for two samples which include both paired observations and independent observations. Journal of Modern Applied Statistical Methods. 16(1), 137-157.

Derrick, B., Toher, D. & White, P. (2016). Why Welch's test is Type I error robust. The Quantitative Methods for Psychology. 12(1), 30-38.

Derrick, B., Toher, D. & White, P. (2017). How to compare the means of two samples that include paired observations and independent observations: A companion to Derrick, Russ, Toher and White (2017). The Quantitative Methods for Psychology. 13(2), 120-126.

Derrick, B. and White, P. (2017) Comparing two samples from an individual Likert question. International Journal of Mathematics and Statistics. 18(3), 1-13.

Derrick, B., White, P., and Toher, D.

Paley, R. E. A. C. & Wiener, N. (1934). Fourier transforms in the complex domain. American Mathematical Society. 19

Sakia, R. M. (1992). The Box-Cox transformation technique: a review. The Statistician. 169-178.

Servin, B. & Stephens, M. (2007). Imputation-based analysis of association studies: candidate regions and quantitative traits. PLoS Genet, 3(7), e114.

Tukey, J. W. (1962). The future of data analysis. The Annals of Mathematical Statistics, 33(1), 1-67.

Van der Waerden, B. L. (1952). Order tests for the two-sample problem and their power. Indagationes Mathematicae, 14(253), 453-458.

Zarembka, P. (1972). Transformation of variables in econometrics. Transformation of variables in econometrics. In Econometrics, 261-264.